\documentstyle[12pt,aaspp4]{article}

\newcommand{\etal}{{\it et al.}}
\newcommand{\chisq}{\mbox{$\chi^{2}$}}
\newcommand{\kms}{km\,s$^{-1}$}

\newcommand{\msun}{\mbox{M$_{\odot}$}}
\newcommand{\dmass}{\mbox{M$_{D}$}}
\newcommand{\mdot}{\mbox{$\dot{M}$}}
\newcommand{\mdotedd}{\mbox{$\dot{M}_{E}$}}
\newcommand{\mdote}{\mbox{$\dot{m}$}}
\newcommand{\mdotc}{\mbox{$\dot{m}_{c}$}}
\newcommand{\smy}{\mbox{M$_{\odot}$\,yr$^{-1}$}}

\newcommand{\muas}{\mbox{$\mu$as}}
\newcommand{\bq}{\begin{equation}}
\newcommand{\eq}{\end{equation}}

\newcommand{\ergs}{\mbox{erg\,s$^{-1}$}}
\newcommand{\rg}{\mbox{R$_{S}$}}
\newcommand{\rtrans}{\mbox{$r_{tr}$}}
\newcommand{\rtt}{\mbox{$r_{22}$}}
\newcommand{\dvsys}{\mbox{$\delta v$}}

\begin{document}
\vskip 0.6 in
\noindent
 
\title{VLBA Continuum Observations of NGC 4258: Constraints
on an Advection-Dominated Accretion Flow}
 
\author{
J. R. Herrnstein\altaffilmark{1},
L. J. Greenhill\altaffilmark{2}, 
J. M. Moran\altaffilmark{2},
P. J. Diamond\altaffilmark{1},
M. Inoue\altaffilmark{3},
N. Nakai\altaffilmark{3}, and
M. Miyoshi\altaffilmark{4},
}
\altaffiltext{1}{National Radio Astronomy Observatory, PO Box O, Socorro, NM 87801}
\altaffiltext{2}{Harvard-Smithsonian Center for Astrophysics, Mail Stop 42, 
60 Garden Street, Cambridge, MA 02138}
\altaffiltext{3}{Nobeyama Radio Observatory, National Astronomical Observatory, 
Minamimaki, Minamisaku, Nagano 384-13, Japan}
\altaffiltext{4}{Mizusawa Astrogeodynamics Observatory, National Astronomical 
Observatory, 2-12 Hoshigaoka, Mizusawa Iwate 023, Japan}

\slugcomment{Revised February 13, 1998} 
\begin{abstract}
We report a $3\sigma$ upper limit 
of $220$\,$\mu$Jy on any 22-GHz continuum emission coincident with the central 
engine in NGC 4258. If NGC 4258 is powered by an advection-dominated accretion 
flow, this radio upper limit implies that the inner advection-dominated flow 
cannot extend significantly beyond $\sim10^{2}$ Schwarzschild radii.
\end{abstract}



\section{Introduction}

NGC 4258 is a weakly active Seyfert 2 galaxy possessing a highly obscured central X-ray 
source with a 2--10 keV luminosity of $4\times10^{40}$ \ergs\  (Makishima \etal\ 1994) 
and nuclear continuum and narrow line emission seen in reflected, polarized optical light 
(Wilkes \etal\ 1995).  The galaxy also harbors one of the first known nuclear megamasers 
(Claussen, Heiligman, \& Lo 
1984), and VLBA observations reveal a nearly edge-on, extremely thin, 
slightly warped Keplerian disk (Watson \& Wallin 1994; 
Greenhill \etal\ 1995; Miyoshi \etal\ 1995; Moran \etal\ 1995; Herrnstein, Greenhill, \&
Moran 1996).  The masers extend from 
0.13 to 0.26 pc (for a distance of 
6.4 Mpc), and the Keplerian rotation curve requires a central binding mass, $M$,
of $3.5\pm0.1\times10^{7}$ \msun\ within 0.13 pc.  The velocity centroid of the disk 
agrees well with the optically-determined systemic velocity of the galaxy, and 
the rotation axis of 
the disk is aligned with the inner portion of large-scale twisted jets seen in 
radio to X-ray emission (Cecil \etal\ 1992).  VLBA continuum 
observations also reveal a subparsec-scale jet oriented along the disk axis 
(Herrnstein \etal\ 1997).  The central mass estimate provided by the 
maser observations corresponds to an Eddington luminosity, 
$L_{E}$, of 
$4.4\pm0.1\times10^{45}$\,\ergs. The bolometric luminosity, $L$, is more 
difficult to estimate because the central edge-on disk obscures 
the nuclear emission.  Wilkes \etal\ (1995) estimate $L \sim 10^{42-44}$ \ergs\ 
based on observations of the nuclear continuum in optical, polarized light. However, 
the 2--10 keV X-ray flux suggests $L\sim4\times10^{41}$ \ergs, based on the argument 
that the X-ray luminosity typically accounts for $\sim10$\% of $L$ in AGN (Mushotsky, 
Done, \& Pounds 1993).  Hence $L\sim10^{42\pm1}=10^{-3.6\pm1}L_{E}$, and the central 
source in NGC 4258 is highly sub-Eddington.

Between 0.13 and 0.26 pc ($4\times10^{4}$ and $8\times10^{4}$ Schwarzschild radii, \rg), 
the masers appear to trace a cool, thin accretion disk: the temperature in the maser 
layer must be between approximately 300 and 1000\,K to support maser action, and the 
aspect ratio of the layer is less than 0.3\% (Moran \etal\ 1995).  Unfortunately, 
the structure within 0.13 pc is not directly observable, and the precise nature of 
the NGC 4258 central engine remains obscure. One possibility is that the outer thin disk 
traced by the masers extends to the central black hole, and that the NGC 4258 central 
engine is fueled by an optically thick, geometrically thin accretion disk 
(Shakura \& Sunyaev 1973). Since the radiative efficiency, $\eta$ (defined through $L=\eta\mdot 
c^{2}$, where \mdot\ is the accretion rate), of such a disk is high, the sub-Eddington 
luminosity of NGC 4258 then implies a correspondingly sub-Eddington \mdot.  Specifically, 
for $\eta=0.1$, $\mdote\equiv\mdot/\mdotedd=10^{-3.6\pm1}$. Here \mdotedd\ is the 
Eddington accretion rate, given by $2.2\times10^{-8}M$\,\smy.  Neufeld \& Maloney 
(1995) argue that for $\mdote\sim10^{-4.0}\alpha$, the outer disk, which is obliquely 
irradiated by the central X-ray source as a result of the warp, changes from cool 
molecular gas to warm atomic gas at about 0.23 pc, providing a natural explanation 
for the observed outer edge to the maser emission. Here, $\alpha$ is the standard 
Shakura-Sunyaev parameterization of the kinematic viscosity.  In the Neufeld \& 
Maloney model, $\eta\sim10^{-0.6\pm1}\alpha^{-1}$.

Alternatively, Lasota \etal\ (1996; hereafter L96) have proposed that NGC 4258 
harbors an optically thin advection-dominated accretion flow (ADAF) at small radii, 
within the cool molecular disk.  Narayan \& Yi (1994, 
1995a\&b; hereafter NY95b) have demonstrated the stability and self-consistency
of a geometrically thick, optically thin accretion flow in which an extremely 
hot ($T\sim10^{12}$\,K) ion plasma coexists with much cooler ($T\sim10^{9.5}$
\,K) electrons. For the ions, the radiative timescale is much longer than the
accretion timescale. Thus, in the case of accretion onto a black hole (BH), 
the majority of the viscously dissipated energy is carried 
through the event horizon of the BH by the hot ions, and the ADAF radiative 
efficiency is several orders of magnitude less than that of the standard 
thin disk. The appeal of the two-temperature ADAF models is that the 
optically thin synchrotron, 
bremsstrahlung, and Compton emission from the $\sim10^{9.5}$\,K electrons can  
account for the spectra of a broad variety of accreting systems 
over many decades in energy (see Narayan 1997 for a recent review).  For example, 
efforts to model the radio-to-$\gamma$-ray spectrum of the galactic source 
Sgr A$^{*}$ with an
ADAF spectrum are reasonably successful (Narayan, Yi, \& Mahadevan 1995; 
Manmoto, Mineshige, \& Kusunose 1998; Narayan \etal\ 1998).  This is an 
important result: optically 
thick, thin disk models emit largely as blackbodies and are generally 
too cool to account for the X-ray emission observed in accreting binaries and 
active galactic nuclei (AGN).

The two-temperature ADAF solution is valid for accretion rates less than 
$\mdotc\sim1.3\alpha^{2}$ (Esin \etal\ 1997) or, equivalently, for bolometric 
luminosities less than $\sim1.3\alpha^{2}L_{E}$.
NGC 4258 satisfies this requirement for $\alpha>10^{-1.9\pm0.5}$. In general, 
the ADAF models are parameterized in terms of $M$, $\alpha$, $\mdote$, 
$\beta$ (the ratio of gas pressure to total pressure), and \rtrans\ (the 
transition radius from the outer thin disk to the inner ADAF).
The NGC 4258 ADAF model of L96 
is constrained primarily by the maser disk binding mass and the 2--10 keV 
X-ray luminosity and spectral index, and secondarily by the polarized optical 
flux.  L96 assume $\beta=0.95$ and find that an NGC 4258 ADAF 
requires $\mdote\sim10^{-1.8}\alpha$.  More recent calculations 
suggest $\mdote\sim10^{-1.6}\alpha$ (Narayan, personal communication), and the 
condition $\mdote\la\mdotc$ requires $\alpha\ga0.02$ in NGC 4258.  In this model, 
$\eta\sim10^{-3\pm1}\alpha^{-1}$, and the NGC 4258 ADAF is very rapid, 
but very inefficient in producing radiation. 

In general, the hot ADAF electrons are expected to generate 
significant radio emission via the thermal synchrotron process.  Unfortunately, 
most of the underluminous AGN and quiescent ellipticals which plausibly harbor 
ADAFs also possess core-jet radio structures that are not directly associated
with the ADAF itself.  In practice it is extremely 
difficult to discriminate this emission from the ADAF synchrotron emission.  Even 
when VLBI provides sufficient angular resolution to resolve individual components 
within the radio core, it is usually impossible to determine which, if any, of 
this emission originates precisely at the central engine, where the ADAF emission 
must arise.  Because the relative position of the center of mass of the sub-parsec 
molecular disk can be measured to a fraction of a milliarcsecond with VLBI, 
NGC 4258 provides a rare opportunity to circumvent the ambiguity associated 
with core-jet emission, and to test the radio portion of the proposed ADAF 
spectrum directly.  

\section{Observations and Analysis}

NGC 4258 was observed at 22 GHz on 1997 March 7, March 23, and April 7 for 12, 14, 
and 15 hours, respectively, with the VLBA, phased VLA, and 140-foot Green Bank telescope 
of the NRAO 
\footnote{The National Radio Astronomy Observatory is operated by Associated 
Universities, Inc, under cooperative agreement with the National Science Foundation.}, 
and the 100-meter MPIfR telescope.
Eight, 8-MHz bands were observed, two of which were dedicated to observing the 
systemic maser emission centered at 470 
\kms\ (LSR, radio definition). The maser emission in NGC 4258 extends from 
$\sim-500$\,\kms\ to 1500\,\kms, and the remaining bands were positioned 
between 1707 and 2354\,\kms\ and $-738$ and $-1386$\,\kms\ to provide 
continuum coverage devoid of maser 
emission.  In each epoch, 14-minute scans 
of NGC 4258 were interleaved with 3-minute observations of a nearby delay 
and fringe-rate calibrator (either 1308+33 or 4C39.25).  
The VLA was phased prior to each scan with
observations of either 1144+399 or 1308+33.  

Calibration was performed in AIPS using standard techniques. Gross 
amplitude calibration was accomplished using recorded 
system temperatures and tabulated values for the antenna system
equivalent flux densities at 22 GHz.  Amplitude
self-calibration was performed on a strong maser feature, and we estimate 
the absolute amplitude calibration to be accurate to about 10\%, or about 
0.2 mJy for 
the NGC 4258 continuum flux.  This dominates the 0.1\,mJy thermal noise 
in the continuum images. Phase 
self-calibration was performed on the strong maser at 492\,\kms\ 
after correcting 
for any residual clock offsets and drifts, and the resulting phase corrections 
were 
applied to the entire dataset in order to stabilize the interferometer.  

\begin{figure}[pht]     
\plotone{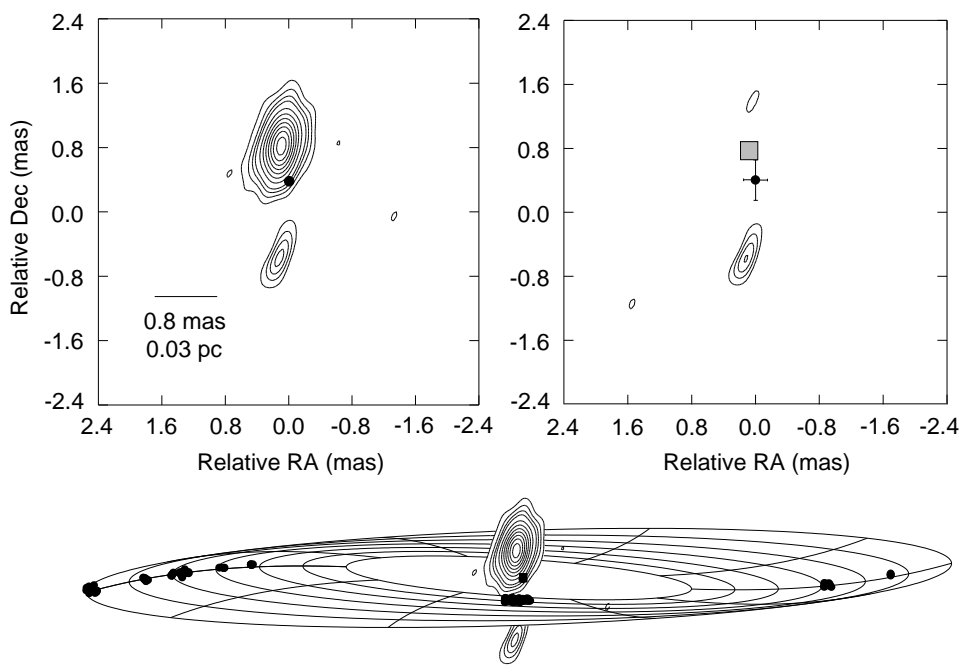}
\caption {{\it Upper left: }22-GHz VLBA continuum image of NGC 4258. Three,
14-hour observations have been combined in generating the image. The filled circle 
marks the center of mass of the maser disk. The thermal noise in the map is 72 $\mu$Jy.
{\it Upper right: }Continuum image after deconvolving the northern jet emission 
from the ({\it u,v}) dataset. The clean box used in the doconvolution is shown as a 
filled square.  The disk center is marked with 5$\sigma$ uncertainties.
{\it Bottom: }The continuum map superposed on the best-fitting warped disk
model.  The masers are shown as filled circles, the disk center as the filled square.}
\label{fg:ljgc}
\end{figure}

Figure~1 shows the weighted average of the images, as well as the best-fitting 
warped-disk model. The disk model, and in particular the disk center, is determined 
using a global \chisq\ 
minimization code (Herrnstein 1997) in which the 
maser positions and line-of-sight (LOS) velocities are used to define the 
best-fitting 
warped-disk model, parameterized in terms of the disk center, the central 
mass, and 
two polynomials in radius describing the three-dimensional warp of the disk.
Formal 
parameter uncertainties are estimated using a Monte Carlo analysis in which 
\chisq\ minimizations are performed on an ensemble of model disks sampled 
consistently with the actual data.  Herrnstein (1997) reports uncertainties in the fitted 
disk center of 30\,\muas\ ($=1\times10^{-2}$\,pc $=3\times10^{2}$\,\rg) 
and 50\,\muas\ ($=2\times10^{-2}$\,pc $=5\times10^{2}$\,\rg) in right 
ascension
and declination, respectively.

The 22-GHz continuum emission traces a jet-like structure oriented along the 
rotation axis of 
the maser disk and comprised of distinct northern and southern components that 
are each unresolved by the 0.6 by 0.3 mas beam. The flux density of the 
northern component varied from 2.5 to 3.5 ($\pm0.3$) mJy during  
the three epochs, while the southern component remained constant at $0.5\pm0.1$ 
mJy.  These correspond to brightness temperature lower limits of 
$1\times10^{7}$ and $2\times10^{6}$\,K for the northern and southern 
components, respectively.  The centroid of the northern component appeared 
between $0.35\pm0.02$\,mas and $0.46\pm0.01$\,mas 
north of the disk center, while the southern feature remained stationary 
at $-1.0\pm0.06$ mas.  The new data are 
consistent with earlier studies of the NGC 4258 jet (Herrnstein \etal\ 
1997). The $\sim0.4$\,mas ($4\times10^{16}$\,cm or $4\times10^{3}$\,\rg) 
offset of the northern jet emission from the disk center is highly 
significant compared to the $\sim0.05$ mas statistical uncertainty, 
and can be explained by optical depth effects along a mildly relativistic 
jet (Blandford \&  K$\ddot{\mbox{o}}$nigl 1979; Herrnstein \etal\ 1997).  The 
relative faintness of the southern jet is probably due to thermal free-free 
absorption in a thin surface layer of ionized gas along the near edge of the 
disk, which intercepts the LOS to the southern component. In this case, the 
relatively large offset of the southern source from the disk center occurs 
because the disk free-free opacity is approximately proportional to $1/r^{2}$, where $r$ 
is the distance from the central X-ray source (Herrnstein, Greenhill, \& 
Moran 1996).  

We searched for continuum emission at the fitted-disk center of mass in 
each epoch by removing the 
northern jet emission from the ({\it u,v})-data and re-imaging.  This 
subtraction was accomplished by extracting from the ({\it u,v})-data only that emission 
located within a $\sim0.24\times0.24$ mas clean-box centered 
on the peak of the northern jet emission.  In each epoch, this box lay 
north of the $5\sigma$ confidence interval for the disk center, and it is 
extremely unlikely that any central continuum emission has been inadvertently 
removed in the deconvolution process. There is no evidence for any central 
22-GHz continuum emission within the 5$\sigma$ interval around the disk center in 
any of the three epochs.  Figure~1 shows the weighted average of the deconvolved 
images, together with the clean-box (for the second epoch) and the 
$5\sigma$ uncertainty in the position of the disk center. The map has
a thermal noise of about 72\,$\mu$Jy, and {\it we report a 3$\sigma$ upper 
limit of 220 $\mu$Jy on any continuum emission coincident with the central mass
of the NGC 4258 maser disk.}  This corresponds to a brightness temperature lower 
limit of $8\times10^{5}(r_{a}/5\times10^{16}\mbox{ cm})^{-2}$\,K, where $r_{a}$ is the 
(undetected) source size, in cm.  The outer thin disk is tipped down from 
the LOS by $8\pm1^{\circ}$ at 0.13 mas, and, given the shape of the warp 
between 0.13 and 0.26 pc, it is unlikely that the thin molecular disk significantly
obscures any central radio emission. 

\section{Discussion}
\label{s:adaf.3}

The most direct method to discriminate between the standard thin-disk and ADAF 
accretion modes would be to determine the accretion rate through the outer thin disk.  
For a Shakura-Sunyaev accretion disk in hydrostatic equilibrium (c.f. Frank, King, \& 
Raine 1992)
\bq
\frac{\mdote}{\alpha}\simeq 530\frac{\dmass c_{s}^{2}}{M^{3/2}(r_{2}^{1/2}-r_{1}^{1/2})} ~~,
\label{eq:danger}
\eq
where \dmass\ is the disk mass (in solar masses) between $r_{1}$ and $r_{2}$ (in pc), 
and $c_{s}$ is the isothermal sound speed in \kms. 
The observed upper limit in the thickness of the maser layer, together 
with the mere presence of the masers, implies $c_{s}\la2.5$\,\kms\  (or $T \la 1000$K)
(Moran \etal\ 1995).  The simplest assumption is that the masers lie in the midplane of the disk 
and that the midplane density must be less than $\sim10^{10}$\,cm$^{-3}$ to avoid 
thermal quenching of the population inversion. In this case, $\dmass\la5.8\times10^{4}$\,\msun\ 
between 0.13 and 0.26 pc and $\mdote/\alpha\la10^{-2.2}$.  A second, more
observationally motivated constraint on \dmass\, comes from the precision of the 
Keplerian rotation curve of the high-velocity masers.  While the statistical scatter 
in the LOS velocities around the best-fitting Keplerian rotation curve is about 3\,\kms, 
the systematic deviation from Keplerian rotation (\dvsys) across the masing 
zone may be considerably larger due to uncertainties in the fitted-disk parameters, 
which are partially correlated.  A formal \chisq\ analysis demonstrates 
$\dvsys\la20$\,\kms.  This requires  $\dmass\la1.5\times10^{6}$\,\msun\ and 
$\mdote/\alpha\la10^{-0.8}$.  This is about six times larger than the ADAF 
accretion rate.  Thus, the maser data do not definitively rule out either mode of 
accretion in NGC 4258, and the present continuum observations provide an important 
additional constraint on the models.

We associate the center of the maser disk with the NGC 4258 central engine, and 
interpret the continuum non-detection in the context of the ADAF models using
the analytical expressions presented in Mahadevan (1997; hereafter M97), which 
duplicate reasonably well the 
numerical results of NY95b.  M97 includes all the relevant electron 
heating and cooling terms, but considers only the region between 3 and $10^{3}$ \rg, 
assumes that the electron temperature, $T_{e}$, is constant and much less than 
the ion temperature over these radii, and treats the flow as purely spherical. These 
approximations are justified by the more detailed analysis of NY95b. 

M97 treats the ADAF as a relativistic thermal plasma, which is valid for 
$\mdote\ga10^{-4}\alpha^{2}$ (Mahadevan \& Quataert 1998) and is justified in 
NGC 4258.  M97 shows that the radius, $r_{\nu}$, at which the spherical 
flow becomes optically thick to synchrotron self-absorption depends on frequency, 
$\nu$, as
\bq
r_{\nu}=0.47T_{e}^{8/5}\nu^{-4/5}M^{-2/5}\mdote^{3/5}\alpha^{-2/5}\mbox{\,   \rg},
\label{eq:maha1}
\eq
where  we have assumed $\beta=0.5$.  To a 
very good approximation, the synchrotron spectrum, $L_{\nu}$, is completely 
determined by the emission at $r_{\nu}$, which may be approximated by a blackbody 
spectrum, $B_{\nu}$, in the Rayleigh-Jeans limit (M97): 
\bq
L_{\nu}=\pi B_{\nu}4\pi r_{\nu}^{2}=2.3\times10^{-25}T_{e}^{21/5}\nu^{2/5}M^{6/5}\mdote^{6/5}\alpha^{-4/5}\mbox{   \ergs}.
\label{eq:maha2}
\eq
Equations~\ref{eq:maha1} and \ref{eq:maha2} show that both $r_{\nu}$ and $L_{\nu}$
are steep functions of the electron temperature.  However, the analytical 
treatment of M97 and the more rigorous numerical work of NY95b both indicate that 
for high accretion-rate systems, the electron temperature is confined to a narrow 
range, and is essentially independent of the central mass. Specifically, M97 finds 
that for $\mdote\sim10^{-1.5}\alpha$, $2.1\times10^{9} \la T_{e} \la 3.1\times10^{9}$ 
over 9 decades in $M$. 

For $\mdote=10^{-1.6}\alpha$, $M=3.5\times10^{7}$\,\msun, and
$T_{e}=2.15\times10^{9}$\,K, equation~\ref{eq:maha2} indicates that 
$F_{22}\ga220$\,$\mu$Jy for $\alpha>0.009$. Here $F_{22}$ is the 22-GHz ADAF flux 
density assuming a distance of 6.4\,Mpc.  For $\alpha\la10^{-2}$, $\mdote\ga\mdotc$ 
and advection is not a viable solution.  Hence, the ADAF models significantly 
overestimate the actual 22-GHz emission over the full range of allowed $\alpha$.  
Here, we have assumed that \rtrans\ is larger than the effective 22-GHz photosphere 
(\rtt) as given by equation~\ref{eq:maha1}, a necessary prerequisite to the assumption 
of constant $T_{e}$.  However, the absence of detectable 22-GHz continuum emission in 
NGC 4258 may be consistent with an ADAF that is truncated such that $\rtrans<\rtt$.  
Equation~\ref{eq:maha1} indicates that $\rtt\simeq300\alpha^{1/5}$\,\rg\ for the NGC 4258
ADAF, suggesting that $\rtrans\la10^{2}$\,\rg\ for typical values of $\alpha$.  A second, 
more stringent upper limit on \rtrans\ comes from assuming that, for those frequencies 
below which the condition $r_{\nu}\ga\rtrans$ is satisfied, the emission follows the 
$T=2.5\times10^{9}$ blackbody spectrum of the outermost ADAF shell at $r=\rtrans$.  
In this case, the 22-GHz upper limit requires $\rtrans\la80$\,\rg.  {\it Thus, the 
220\,$\mu$Jy, 3$\sigma$ upper limit on any compact 22-GHz emission coincident with the 
central engine in NGC 4258 implies that the proposed central ADAF cannot extend 
significantly beyond $\sim10^{2}$\,\rg}. More rigorous numerical 
methods also lead to ADAF solutions consistent with the present 22-GHz upper limit for 
$\rtrans\la10^{2}$\,\rg\ (Narayan, personal communication).

In addition to affecting the radio emission, the transition radius affects both
the hard X-ray spectrum and the optical/UV emission.  In the former case, the 
ADAF is Compton cooled by the soft photons from the outer cool disk, and as \rtrans\ 
moves inward, the hard X-rays are suppressed.  L96 find that $\rtrans\ga10$\,\rg\ 
is consistent with the Makishima \etal\ (1994) measurements.  The optical/UV
spectrum depends on \rtrans\ because the  blackbody spectrum of the non-molecular 
portion of the disk peaks at 
these wavelengths.  Wilkes \etal\ (1995) have measured the 5500 \AA\ 
flux  toward NGC 4258 in polarized light. This emission presumably arises in 
the central engine and has been scattered into our LOS. The corresponding 
5500 \AA\ central engine luminosity is highly sensitive to the type of scattering 
screen invoked (dust or electrons), and Wilkes \etal\ (1995) estimate that 
$L_{5500}\sim10^{37-39}$\,\ergs\,\AA$^{-1}$.  L96 find that the NGC 4258 ADAF contributes 
only about $10^{36.3}$\,\ergs\,$\AA^{-1}$\ at 5500\,\AA, and here we estimate 
the contribution of the thin disk to the 5500\,\AA\ luminosity for a range 
of \rtrans.

The inferred isotropic luminosity of an optically thick, steady, thin disk 
truncated at \rtrans\ is (c.f. Frank, King, \& Raine 1992):
\bq
L_{\lambda}^{(td)}\simeq\frac{16\pi^{2}h\nu^{2}\cos{i}}{\lambda^{3}}\int_{\rtrans}^{\mbox{$r_{out}$}}\frac{rdr}{e^{h\nu/kT(r)}-1}\mbox{  \ergs\,\AA$^{-1}$},
\label{eq:fkr1}
\eq
where $T(r)=2.2\times10^{5}\mdot_{26}^{1/4}M_{8}^{1/4}r_{14}^{-3/4}$ K, $i$ 
is the angle between the disk normal and the LOS ($\sim82^{\circ}$ in NGC 4258), 
$\mdot_{26}$ is the accretion rate in units of $10^{26}$\,g\,s$^{-1}$, 
$M_{8}$ is the central mass in units of $10^{8}$\,\msun, and $r_{14}$ is the
radius in units of $10^{14}$\,cm.  For $\mdot/\alpha=10^{-1.7}$\,\smy\ and 
$r_{out}=4\times10^{4}$\,\rg, equation~\ref{eq:fkr1} indicates that 
$L_{5500}^{(td)} \la 10^{37}$\,\ergs\,\AA$^{-1}$ for $\rtrans\ga10^{2}$\,\rg. 
{\it At the accretion rates required by the ADAF models, the outer thin disk 
must extend to within $\sim10^{2}$\,\rg\ in order to generate sufficient optical 
luminosity.  This is additional, independent evidence that 
$\rtrans\la10^{2}$\,\rg\ in NGC 4258.}  

The masers in NGC 4258 extend from 0.13 pc to 0.26 pc, and the cause of the
observed inner edge at $4\times10^{4}$\,\rg\ is 
uncertain.  The present observations effectively disqualify one of the most
promising theories: that the inner edge of the masing zone corresponds to 
\rtrans\ and to the inner edge of the outer cool disk itself. Even if NGC 4258
is powered by an ADAF, the lack of 22-GHz emission requires that the outer 
thin disk extends to well within $4\times10^{4}$\,\rg, and another explanation 
is needed to account for the lack of masers in this region.  We note that 
Neufeld \& Maloney (1995) argue that a proposed flattening of the disk at 
small radii would reduce the disk's exposure to central, hard X-rays, and may
account for the lack of maser emission within 0.13\,pc.

There are a great many low-luminosity accreting systems, encompassing a wide 
variety of types and spanning many decades in central mass, that are apparently 
sufficiently sub-Eddington to be plausible ADAF candidates.  However, for most of 
these systems, a standard Shakura-Sunyaev thin disk is also a viable solution.  
Thus, while there is compelling evidence for the existence of ADAFs in 
some systems, it remains unclear to what extent the optically thin, two-temperature 
ADAF is in general applicable.  Unfortunately, a physically motivated explanation 
for how ADAFs are triggered in the first place 
remains elusive, and, until this `on-switch' is discovered, questions concerning 
the universality of ADAFs must necessarily be addressed empirically. NGC 4258 
is a useful place to explore this issue, since it is highly sub-Eddington, and 
because the nuclear maser and VLBI together provide access to a host of usually 
obscure parameters.  While the present 22-GHz non-detection does not rule out the 
presence of an ADAF in NGC 4258, it does begin to place interesting constraints on 
the geometry of any proposed ADAF. 

\acknowledgments{We thank R. Mahadevan and R. Narayan for helpful discussions, and C.
Henkel for valuable assistance in conducting the observations at Effelsberg.}

\end{document}